\documentclass[sigconf]{acmart}

\renewcommand\footnotetextcopyrightpermission[1]{}

\AtBeginDocument{%
  }

\acmConference[]{}
\acmBooktitle{}
\acmDOI{}
\acmISBN{}

\usepackage{listings}
\usepackage{cleveref}
\usepackage{xcolor}

\lstdefinelanguage{diff}{
  morecomment=[f][\color{red}]{-},
  morecomment=[f][\color{blue}]{+},
  morecomment=[f][\color{gray}]{@@},
  morecomment=[f][\color{purple}]{---},
  morecomment=[f][\color{purple}]{+++},
}

\begin{document}

\title{Reducing Token Usage of State-in-Context Agents using Minification}

\author{Nicolas Hrubec}
\email{nicolas.hrubec@tuwien.ac.at}
\orcid{0009-0002-8831-8314}
\affiliation{%
  \institution{TU Wien}
  \city{Vienna}
  \country{Austria}
}

\author{Jürgen Cito}
\email{juergen.cito@tuwien.ac.at}
\orcid{0000-0001-8619-1271}
\affiliation{%
  \institution{TU Wien}
  \city{Vienna}
  \country{Austria}
}

\renewcommand{\shortauthors}{Nicolas Hrubec and Jürgen Cito}

\begin{abstract}
This paper presents a replication and extension of the recently introduced \emph{state-in-context} agent framework. We independently re-implement the \emph{DirectSolve} variant and evaluate it on the SWE-bench Verified benchmark. We report end-to-end full-benchmark results using GPT-5-mini and run selected ablations with GPT-4.1. In addition, we investigate a complementary research question: What is the impact of token-reducing input transformation strategies on the performance of software engineering agents? Based on a preliminary prompt analysis, we identify source code as the dominant contributor to token consumption. We therefore apply a series of code minification techniques that remove or shorten non-essential lexical elements while preserving program semantics. The proposed transformations are integrated into the agent and systematically evaluated. Experiments show that minification reduces average input token usage by 42\% with a 12 percentage-point drop in resolution rate. These findings demonstrate that lightweight source code transformations can yield substantial efficiency gains while retaining a substantial fraction of the baseline performance, indicating a promising path toward more cost-effective agents. The full implementation is publicly available on GitHub\footnote{\url{https://github.com/ipa-lab/minified-state-in-context-agent}}.
\end{abstract}



\maketitle

\fancyhead[]{}

\section{Introduction}

Large language models (LLMs) are increasingly deployed as the central reasoning component in agentic systems for software engineering (SE). On repository-level tasks such as SWE-bench~\cite{jimenez2024swebench}, agents repeatedly retrieve, read, and reason over large codebases. In such workflows, context size grows rapidly, resulting in high token usage. Since token usage typically determines LLM cost, long contexts also incur high operational cost. Additionally, studies have shown that LLMs can struggle to fully utilize long contexts, leading to performance degradation~\cite{hong2025contextrot, liu2023lostmiddlelanguagemodels, laban2025llmslostmultiturnconversation, vishwanath2025medicallargelanguagemodels, shi2023largelanguagemodelseasilydistracted}. Thus, effectively reducing context size (while retaining relevant information) is essential for both performance and cost efficiency in LLM-based SE agents.

Jiang et al.~\cite{jiang2025puttingcontextsimplifyingagents} recently proposed \emph{state-in-context} agents as a way to simplify the architecture of such systems. Rather than reconstructing task-relevant state incrementally through tool calls, a state-in-context agent receives (a compressed version of) the entire environment state in a single prompt and is asked to produce a fix directly. In their \emph{DirectSolve} variant, the LLM first ranks repository files by estimated relevance to the issue description, then aggregates top-ranked files into a large repair context, and finally generates a patch in a SEARCH/REPLACE format. This design removes much of the handcrafted scaffolding and explicit tool interaction present in previous agents such as SWE-Agent~\cite{yang2024sweagentagentcomputerinterfacesenable} or Agentless~\cite{xia2024agentlessdemystifyingllmbasedsoftware}, while maintaining similar performance on SWE-bench Verified~\cite{chowdhury2024swebenchverified}.

\begin{figure}
    \centering
    \includegraphics[width=\columnwidth]{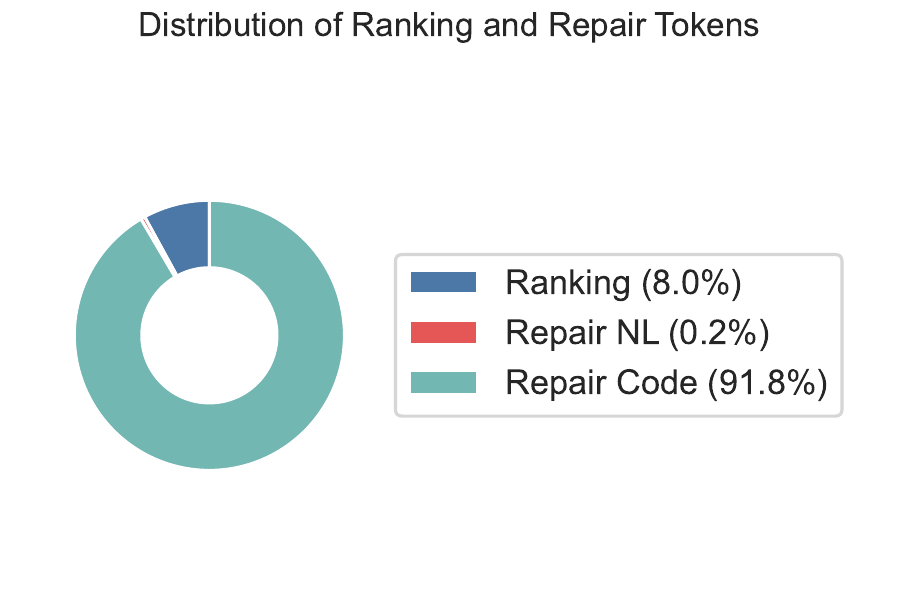}
    \caption{Distribution of Ranking and Repair Tokens.}
    \label{fig:ranking-repair-code-nl}
\end{figure}

This paper addresses reducing end-to-end token consumption in SE agents while aiming to limit performance impact. To make progress toward this goal, we study \emph{state-in-context agents} in greater detail. To understand where tokens are spent in this setting, we first perform a prompt analysis through our re-implementation and replication of \emph{DirectSolve}. We find that the overwhelming majority of tokens are source code tokens consumed during the repair step. Even after capping the repair context at 100{,}000 tokens, more than 90\% of all tokens are spent on source code in the repair prompt. In contrast, ranking and natural language instructions together account for less than 10\% (see \Cref{fig:ranking-repair-code-nl}).

In other aspects of software engineering, code size is routinely reduced through \emph{minification}. Minification removes characters that are not required for program execution but aid human readability~\cite{mdn_minification}. Typical operations include removing whitespace, comments, and line breaks, merging adjacent statements, and renaming identifiers. In the JavaScript ecosystem, for instance, minification tools are standard components of build pipelines to reduce bundle size and improve load times. Motivated by this, our solution focuses on reducing the code size of the state-in-context repair prompt using minification-style code transformations. These transformations are applied to the retrieved source files before they are passed to the LLM for repair. The overall objective is to reduce token consumption in the repair step while maintaining the model’s ability to reason about the underlying program logic. 

\paragraph{Replication} This paper constitutes an independent replication and extension of the DirectSolve state-in-context agent introduced by Jiang et al.~\cite{jiang2025puttingcontextsimplifyingagents}. 
Since DirectSolve is not accompanied by a publicly available code artifact, our replication is based on a clean-room re-implementation derived from the specifications and experimental details provided in Jiang et al.’s paper~\cite{jiang2025puttingcontextsimplifyingagents}.
Minor implementation deviations aimed primarily at reducing overall cost are outlined in \Cref{sec:agent-implementation}. Following the original paper, we evaluate on SWE-bench Verified~\cite{jimenez2024swebench}, with the main difference in our base setup being the choice of underlying LLMs. Our evaluation on SWE-bench Verified is conducted using GPT-5-mini~\cite{openai2025gpt5} for end-to-end results on the full benchmark. We additionally use GPT-4.1~\cite{openai2024gpt4technicalreport} for selected ablation analyses. With this setup, we replicate the main qualitative finding of the original work that a single-call, tool-free state-in-context agent can achieve competitive performance on SWE-bench Verified when paired with a sufficiently capable model. Using GPT-5-mini, we obtain a 50\% pass@1 rate consistent with the originally reported 50.8\% pass@1 result for \emph{DirectSolve} with Gemini-2.5-Pro~\cite{comanici2025gemini25pushingfrontier}. However, we do not attempt to replicate any of the original paper’s ablations on individual agent design choices and file localization errors. 

\paragraph{Extension}
Beyond replication, we extend the original study by systematically investigating source code minification as an explicit context-reduction mechanism for state-in-context agents. While Jiang et al.~\cite{jiang2025puttingcontextsimplifyingagents} employ compression primarily as a preprocessing step to fit long contexts, we treat code transformation itself as a first-class experimental variable. We design and evaluate a set of lightweight, semantics-preserving minification techniques tailored to Python source code, integrate them into the repair pipeline and analyze their individual and combined effects on token usage and repair performance. This extension enables a more fine-grained understanding of how different classes of code-level transformations trade off efficiency and effectiveness in large-context software engineering agents.

\paragraph{Contributions} In summary, we make the following contributions:

\begin{itemize}
\item We re-implement the DirectSolve state-in-context agent of Jiang et al.~\cite{jiang2025puttingcontextsimplifyingagents} and evaluate it on SWE-bench Verified, demonstrating that our implementation reproduces the reported performance trends.
\item We conduct an analysis of token allocation in this pipeline, revealing that source code in the repair prompt dominates token usage and that natural-language instructions and ranking prompts are comparatively negligible.
\item We propose a set of lightweight, semantics-preserving code minification transformations tailored to Python and integrate them into the state-in-context repair workflow.
\item We empirically evaluate these transformations on SWE-bench Verified and find that minification can reduce average input token usage during repair by approximately 42\% with a 12\% absolute drop in resolution rate, leading to significant cost savings while retaining a substantial fraction of the baseline performance. Further analysis indicates that these trends remain largely stable across issue difficulty levels and repositories.
\item We perform ablation studies isolating individual transformations and stacked combinations, quantifying their respective token savings and performance impacts. We additionally conduct a high-level failure analysis to characterize the most common error modes introduced by these transformations.
\end{itemize}

\section{Related Work}

While there is an abundance of literature on agentic systems, research focusing on context management as a primary research target is relatively sparse. Despite its impact on performance and cost, most existing work treats context management as an implementation detail. 

MEM1~\cite{zhou2025mem1} introduces a reinforcement learning framework for updating a compact internal memory state in multi-hop QA~\cite{hotpotqa} and web navigation tasks~\cite{web-navigation}. Instead of appending the entire interaction history to the prompt, MEM1 maintains a compressed shared representation that integrates new observations while discarding redundant information, allowing the agent to operate with constant memory. In contrast, Lindenbauer et al.~\cite{lindenbauer2025complexitytrapsimpleobservation} focus specifically on software engineering agents. Within the SWE-Agent framework~\cite{yang2024sweagentagentcomputerinterfacesenable}, they compare summarization-based and masking-based strategies for managing agent context. Their experiments show that simple observation masking can outperform summarization in efficiency while maintaining comparable performance. Another recent approach by Xiao et al.~\cite{agentdiet} introduces AgentDiet, a trajectory-reduction method that uses an LLM to identify and prune useless, redundant, or expired segments of an agent's history. Their experiments report significant input token reductions while maintaining the same agent performance.

Recent work has also examined the cost implications of code readability and formatting for LLMs, showing that removing non-semantic formatting elements can substantially reduce token usage with limited impact on code-generation tasks~\cite{pan2025hiddencostreadabilitycode}. In contrast, our work studies code minification as a systematic context-reduction mechanism within an end-to-end, repository-level software engineering agent and evaluates its effect on automated bug fixing.

\section{Method}

\subsection{State-in-Context Agent Implementation}\label{sec:agent-implementation}

We adopt the \emph{DirectSolve} \emph{state-in-context} agent~\cite{jiang2025puttingcontextsimplifyingagents} as baseline. Since no public implementation is available, we re-implement the agent from scratch. The workflow consists of the following three steps: 

\begin{enumerate}
    \item An LLM is asked to rank repository files by estimated relevance to the issue description, based solely on the file paths. 
    \item Files are aggregated in the context in ranked order until a predefined token budget is reached.
    \item Given relevant source files, an LLM is asked to produce a patch in a SEARCH/REPLACE format.
\end{enumerate}

Our implementation deviates from the original setup in several ways, primarily to reduce cost and simplify the pipeline. First, we represent the repository using a trie-style, depth-indented directory view, similar to the output of the Linux \texttt{tree} command. This reduces redundancy in the ranking prompt by printing common prefixes (folders) only once. In the original paper, the exact repository representation for the ranking step is not further specified. Second, we explicitly instruct the model to rank files only, i.e. to exclude folders from the ranking output. If the returned ranking omits any files present in the repository, we do not complete it. Third, whereas the original paper exhausts the full context window of the underlying LLM, we cap the aggregated code context at a fixed token budget during repair, to control token usage and to enable a fair comparison across models with different context limits. Finally, we do not perform majority voting over multiple samples. Instead, we generate a single patch and allow up to five resampling attempts if the model produces an invalid or incorrectly formatted patch.

\subsection{Minification}

To reduce token usage during repair, we implement minification strategies that remove or shorten non-essential lexical and syntactic elements while preserving program semantics. We rely on \texttt{pyminifier3} and extend it where necessary. As a running example, we use the following snippet from the \texttt{astropy}\footnote{https://github.com/astropy/astropy/tree/main} package (\texttt{astropy/astropy/table/table.py}):

\begin{lstlisting}[language=Python, caption={Original code snippet from astropy}, label={lst:astropy-original}]
def __get__(self, instance, owner_cls):
    """Get the attribute."""
    # For getting from class not an instance
    if instance is None:
        return self

    # If not already stored on `instance`, make a copy of the class
    # descriptor object and put it onto the instance.
    value = instance.__dict__.get(self.name)
    if value is None:
        value = deepcopy(self)
        instance.__dict__[self.name] = value

    # We set _instance_ref on every call, since if one makes copies of instances, 
    # this attribute will be copied as well, which will lose the reference.
    value._instance_ref = weakref.ref(instance)
    return value
\end{lstlisting}

\subsubsection{Line/Operator Whitespace Removal}

This transformation eliminates blank lines, as they contain no information. Additionally, whitespace around operators can be safely removed in Python, e.g., a line "a + b" can be reduced to "a+b". Applied to \Cref{lst:astropy-original}, these transformations yield the following minified snippet:

\begin{lstlisting}[language=Python, caption={After line and operator whitespace removal}, label={lst:whitespace}]
def __get__(self,instance,owner_cls):
    """Get the attribute."""
    # For getting from class not an instance
    if instance is None:
        return self
    # If not already stored on `instance`, make a copy of the class
    # descriptor object and put it onto the instance.
    value=instance.__dict__.get(self.name)
    if value is None:
        value=deepcopy(self)
        instance.__dict__[self.name]=value
    # We set _instance_ref on every call, since if one makes copies of
    # instances, this attribute will be copied as well, which will lose the
    # reference.
    value._instance_ref=weakref.ref(instance)
    return value
\end{lstlisting}

\subsubsection{Comment/Docstring Removal}

We additionally allow removing comments and docstrings. Comments in Python are prefixed by "\#" and are commonly used to explain code, note assumptions or leave reminders for other developers. Docstrings are triple-quoted string literals used for formal documentation. Both can, in theory, be safely removed, as they are ignored at runtime. However, we might lose relevant information for an LLM agent working on the code. Removing docstrings and comments from \Cref{lst:whitespace} results in the following further minified code:

\begin{lstlisting}[language=Python, caption={After comment and docstring removal}, label={lst:comment}]
def __get__(self,instance,owner_cls):
    if instance is None:
        return self
    value=instance.__dict__.get(self.name)
    if value is None:
        value=deepcopy(self)
        instance.__dict__[self.name]=value
    value._instance_ref=weakref.ref(instance)
    return value
\end{lstlisting}

\subsubsection{Dedent}

In contrast to other languages, Python uses indentation instead of braces to define block structure (scopes of if, for, while, def, etc.). This means that indentation carries semantic meaning and cannot be trivially removed. However, Python's parser does not require a specific number of spaces. Instead, what matters is consistency within a block. In practice, many repositories use four spaces, as recommended by PEP8\footnote{\url{https://peps.python.org/pep-0008/}} and, as a result, enforced by various formatting tools. All of this should not matter to an agent, so we can reduce indentation (e.g., using only two spaces per level) before handing the source code to the agent for processing, provided this is done consistently to preserve semantics.

A critical implementation detail is that, without additional guardrails, the LLM produces invalid patches when this transformation is applied, because the patch indentation style does not align with the original code. We attempt to solve this problem by instructing the LLM to output the SEARCH and REPLACE blocks with a four-space indentation style regardless of how the input source files are formatted:

\begin{lstlisting}[language={},caption={}]
Always output the SEARCH and REPLACE blocks with a four-space indentation style, even if you receive source files that use less indentation
\end{lstlisting}

Applying dedentation to \Cref{lst:comment}, yields the following snippet:

\begin{lstlisting}[language=Python, caption={After dedent transformation}, label={lst:dedent}]
def __get__(self,instance,owner_cls):
 if instance is None:
  return self
 value=instance.__dict__.get(self.name)
 if value is None:
  value=deepcopy(self)
  instance.__dict__[self.name]=value
 value._instance_ref=weakref.ref(instance)
 return value
\end{lstlisting}

\subsubsection{Short Identifiers}

This transformation replaces identifier names (variable, function and class names) with shorter aliases.

\paragraph{Non-semantics preserving}

In this strategy, we allow identifiers to be replaced with shorter aliases without providing a mapping to the model. Although this approach maximizes token savings, it may impair reasoning by removing semantic cues conveyed by descriptive variable names, which are crucial for humans to understand the code's intended behavior. Applied to \Cref{lst:dedent}, we get the following updated snippet:
\begin{lstlisting}[language=Python, caption={After non-semantics preserving short identifier replacement}, label={lst:short-nonsemantic}]
def g(self,i,o):
 if i is None:
  return self
 v=i.__dict__.get(self.n)
 if v is None:
  v=deepcopy(self)
  i.__dict__[self.n]=v
 v._r=weakref.ref(i)
 return v
\end{lstlisting}

\paragraph{Semantics preserving with in-context lookup}

To mitigate the loss of semantics, we also implement a lookup-based variant:

\begin{enumerate}
    \item Find identifiers that can be renamed (variables, functions, classes).
    \item For each identifier, we determine whether it is beneficial to replace it, i.e., whether the savings of the replacement offset the token overhead of adding an entry to the mapping. This depends on how frequently the identifier occurs in the source file and on its length.
    \item If replacement is beneficial, we rename all occurrences of the identifier and add an entry to a mapping table.
\end{enumerate}

The constructed mapping table is then prepended to the repair prompt. This allows the model to reconstruct the original names by consulting the table. We add the following instructions to the repair prompt, explaining how to use the mapping to recover the original code:

\begin{lstlisting}[language={},caption={}]
SOURCE MAPS (for shortened identifiers):

The following mappings show the relationship between shortened identifiers and their original names:

<transformation_name>:
<shortened> -> <original>

Use the original names when you output the search and replace blocks. For instance, if the source maps contain an entry '- a -> b', then a is the shortened name that you will find in the code and b is the original. In this case, use b instead of a when you output the search and replace block.
\end{lstlisting}

For the minified code in \Cref{lst:short-nonsemantic}, we might add the following mappings:

\begin{lstlisting}[language=Python, caption={Lookup table example for semantics preserving short identifier replacement}, label={lst:short-semantic}]
# - i -> instance
# - v -> value
\end{lstlisting}

\subsubsection{Merge/Remove Imports}

This transformation has two variations. When merging imports, all imports from our sources are consolidated into a single synthetic file, with duplicates removed. This newly created file, containing merged imports, is provided to the LLM as the first context file during repair. In contrast, remove imports deletes all imports completely. If multiple transformations are applied, this one should be the last, since it breaks AST-parsing, which some other transformations rely on. Because the running example contains no imports, it does not change as a result of this transformation.

\subsection{Integrating Minification in State-in-Context Agents}

We apply transformations before constructing the repair prompt. When identifier renaming or dedentation is active, additional instructions are added to guide the LLM in generating syntactically valid patches. After patch generation, identifier renaming is reversed in both SEARCH and REPLACE blocks using the mapping table when applicable.

To apply patches, we implement a matching procedure that tolerates differences in whitespace, comments, and docstrings, enabling alignment between minified search blocks and original source files. We do not reverse formatting transformations in the REPLACE block, except for identifier mappings. After applying a patch, we extract the diff using \verb|git diff| and restore the original repository state.
This integration allows the agent to operate on reduced inputs while producing valid patches for the original repository code.

\section{Experiments}

In this section, we detail our experimental setup and analyze the results. 

\subsection{Experimental Setup}

\subsubsection{Benchmark}

All experiments are conducted on the SWE-bench Verified benchmark~\cite{chowdhury2024swebenchverified}, a human-filtered high-quality subset of the original SWE-bench benchmark~\cite{jimenez2024swebench}. Each instance in SWE-bench corresponds to a real GitHub issue from one of twelve open-source Python projects. The input to the agent consists of three components: (1) the issue description, which outlines the reported bug or requested feature; (2) the full source repository at the commit preceding the human fix; and (3) the associated test suite, which includes both failing and passing tests. The agent’s objective is to generate a patch that modifies the repository so that all previously failing tests specific to the issue pass, while avoiding regressions. Following Jiang et al.~\cite{jiang2025puttingcontextsimplifyingagents}, we perform ablations on a subset of 100 randomly selected instances.

\subsubsection{LLMs}

We evaluate two LLMs released by OpenAI: GPT-5-mini~\cite{openai2025gpt5} and GPT-4.1~\cite{openai2024gpt4technicalreport}. End-to-end evaluations on the full SWE-bench Verified benchmark are conducted using GPT-5-mini. Single-transformation ablation experiments (Section~\ref{ablation:single-transformations}) are performed using GPT-4.1, while stacked-transformation ablations (Section~\ref{ablation:multiple-transformations}) are evaluated using both models to assess the robustness of observed trends across LLMs. Following Jiang et al.~\cite{jiang2025puttingcontextsimplifyingagents}, we fix the sampling temperature at 0.8 for all experiments.

\subsubsection{Maximum Repair Context Size}

To control repair-stage costs and enable fair comparisons across models with different context limits, we deliberately deviate from the setup of Jiang et al.’s DirectSolve agent, which exhausts the full LLM context window. Instead, we experiment with fixed maximum repair contexts of 20,000 and 100,000 tokens and measure target-file recall. While the 100,000-token maximum maintains a recall of around 91\%, we observe a significant drop to below 70\% for the 20,000-token maximum. Consequently, all repair runs are limited to 100,000 tokens.

\subsubsection{Tokenization}

Token usage is measured using OpenAI’s GPT-4 tokenizer, as implemented in the open-source \texttt{tiktoken} library~\cite{openai-tiktoken}. We use a single tokenizer across all experiments to ensure that token counts are comparable between runs and configurations. Since tokenization can differ across model families, these counts should be interpreted as a consistent proxy for relative context length rather than an exact reflection of the internal tokenization used by each deployed model.

\subsubsection{Preprocessing}

We apply a preprocessing step to exclude file categories unlikely to be relevant for functional repairs. Following Jiang et al.~\cite{jiang2025puttingcontextsimplifyingagents}, we remove all test files, as they primarily serve to verify correctness rather than to contribute repair-relevant code. For similar reasons, we exclude build, benchmark, and documentation files to improve target file recall.

\subsubsection{Metrics}

We report the resolution rate using the official SWE-bench evaluation harness. Additionally, we record input and output token usage, from which we derive the total cost based on the official OpenAI API pricing~\cite{openai-pricing}.

\subsubsection{Hardware}

All repair experiments are executed on a server equipped with 2× AMD EPYC 7702P CPUs (64 cores / 128 threads each) and 512 GB RAM. However, we expect hardware performance to have little impact on overall runtime, as the dominant runtime factor is latency from API calls. Local processing is comparatively lightweight. The server was therefore mainly used for convenience. It should be perfectly feasible to run the agent on any modern laptop. Running the SWE-bench evaluation harness on the server was not possible due to Docker permission issues. Therefore, we set up an e2-standard-4 GCP instance (4 vCPUs, 16 GB memory) to run the evaluation.

\subsubsection{Baseline}

To establish reference points for evaluation, we run all experiments both with and without the proposed token-reduction transformations. The baseline corresponds to an uncompressed setup, allowing us to quantify the effectiveness of minification strategies relative to the raw agent implementation.

\subsection{End-to-End Results on SWE-bench Verified}

\Cref{tab:repair-summary} presents the overall repair performance on SWE-bench Verified using GPT-5-mini. Without minification, the agent achieves a resolution rate of 50.0\% with an average of 90,535 input tokens (standard deviation 16,795). This result is close to the best DirectSolve score reported by Jiang et al.~\cite{jiang2025puttingcontextsimplifyingagents}, where Gemini-2.5 Pro attains 50.8\% pass@1, indicating that our re-implementation reproduces the original method’s effectiveness. Applying minification reduces the average input size to 52,776 tokens (standard deviation 13,697), corresponding to a 42\% reduction in context length. This reduction is accompanied by a 12\% absolute decrease in performance, from 50.0\% to 38.0\% resolved tasks. For performance reasons, we omit the dedent transformation in this run, as it significantly impairs the resolution rate. Overall, the repair experiments indicate that input compression using minification can substantially reduce context size, at the cost of a noticeable reduction in repair effectiveness. As input context size dominates overall cost in the repair stage (we attribute about 93\% of repair cost to input tokens), the observed savings translate to similar cost savings.

\begin{table*}[h!]
\centering
\begin{tabular}{lrrrr}
\hline
Run Type & Resolved & Resolved (\%) & Avg.\ Input Tokens & Std.\ Input Tokens \\
\hline
no-minification  & 252 & 50.0 & 90{,}535 & 16{,}795 \\
minification     & 188 & 38.0 & 52{,}776 & 13{,}697 \\
\hline
\end{tabular}
\caption{Repair results on SWE-bench Verified with GPT-5-mini.}
\label{tab:repair-summary}
\end{table*}

\begin{table*}[h]
\centering
\begin{tabular}{l
                rr
                rr
                r}
\hline
& \multicolumn{2}{c}{no-minification} & \multicolumn{2}{c}{minification} & Token Reduction (\%) \\
Difficulty
& Resolved (\%) & Avg. Input Tokens
& Resolved (\%) & Avg. Input Tokens
& \\
\hline
<15 min fix       & 65.0 & 88{,}921 & 48.0 & 52{,}744 & 40.7 \\
15 min--1 hour    & 45.0 & 91{,}389 & 35.0 & 52{,}248 & 42.8 \\
1--4 hours        & 19.0 & 92{,}721 & 7.0  & 56{,}172 & 39.4 \\
>4 hours          & 0.0  & 90{,}032 & 0.0  & 53{,}320 & 40.8 \\
\hline
\end{tabular}
\caption{Repair results on SWE-bench Verified with GPT-5-mini by instance difficulty.}
\label{tab:repair-summary-by-difficulty}
\end{table*}

\Cref{tab:repair-summary-by-difficulty} presents results grouped by issue difficulty. Difficulty levels correspond to the estimated time engineers reported for resolving each issue, divided into four categories. We consistently reduce token usage by about 40\% with a performance degradation of 10-17\%, indicating that minification effectiveness is largely independent of problem complexity. Finally, \Cref{tab:repair-summary-by-repository} reports the same analysis by repository. Of the 12 repositories in SWE-bench Verified, we exclude those with fewer than 10 samples, resulting in 10 repositories shown in the table. At the repository level we observe larger variations. Token reductions mostly range from 50\% to 70\%. Performance degradation is mostly in the 10–20\% range, with a few pronounced outliers, most notably \texttt{psf/requests}.

\begin{table*}[h]
\centering
\begin{tabular}{l
                rr
                rr
                r}
\hline
& \multicolumn{2}{c}{no-minification} & \multicolumn{2}{c}{minification} & \multicolumn{1}{c}{Token Reduction} \\
Repository
& Resolved (\%) & Avg. Input Tokens
& Resolved (\%) & Avg. Input Tokens
& (\%) \\
\hline
astropy/astropy            & 36.0 & 97632 & 23.0 & 48503 & 50.3 \\
django/django              & 54.0 & 89328 & 43.0 & 56053 & 37.3 \\
matplotlib/matplotlib      & 41.0 & 79768 & 21.0 & 42214 & 47.1 \\
psf/requests               & 75.0 & 70852 & 25.0 & 39990 & 43.5 \\
pydata/xarray              & 45.0 & 84740 & 41.0 & 39305 & 53.6 \\
pylint-dev/pylint          & 30.0 & 94360 & 20.0 & 65980 & 30.1 \\
pytest-dev/pytest          & 63.0 & 97048 & 37.0 & 62709 & 35.4 \\
scikit-learn/scikit-learn  & 72.0 & 96401 & 53.0 & 41613 & 56.8 \\
sphinx-doc/sphinx          & 43.0 & 96235 & 32.0 & 65860 & 31.6 \\
sympy/sympy                & 43.0 & 92840 & 33.0 & 46777 & 49.7 \\
\hline
\end{tabular}
\caption{Repair results on SWE-bench Verified with GPT-5-mini by repository.}
\label{tab:repair-summary-by-repository}
\end{table*}

\subsection{Ablations}

To analyze the contribution of individual transformations, we perform ablation experiments on a subset of 100 benchmark instances. Single-transformation ablations are performed using GPT-4.1, while stacked-transformation ablations are evaluated using both GPT-4.1 and GPT-5-mini.

\subsubsection{Single Transformations}\label{ablation:single-transformations}

\begin{figure*}[t]
    \centering
    \includegraphics[width=0.8\textwidth]{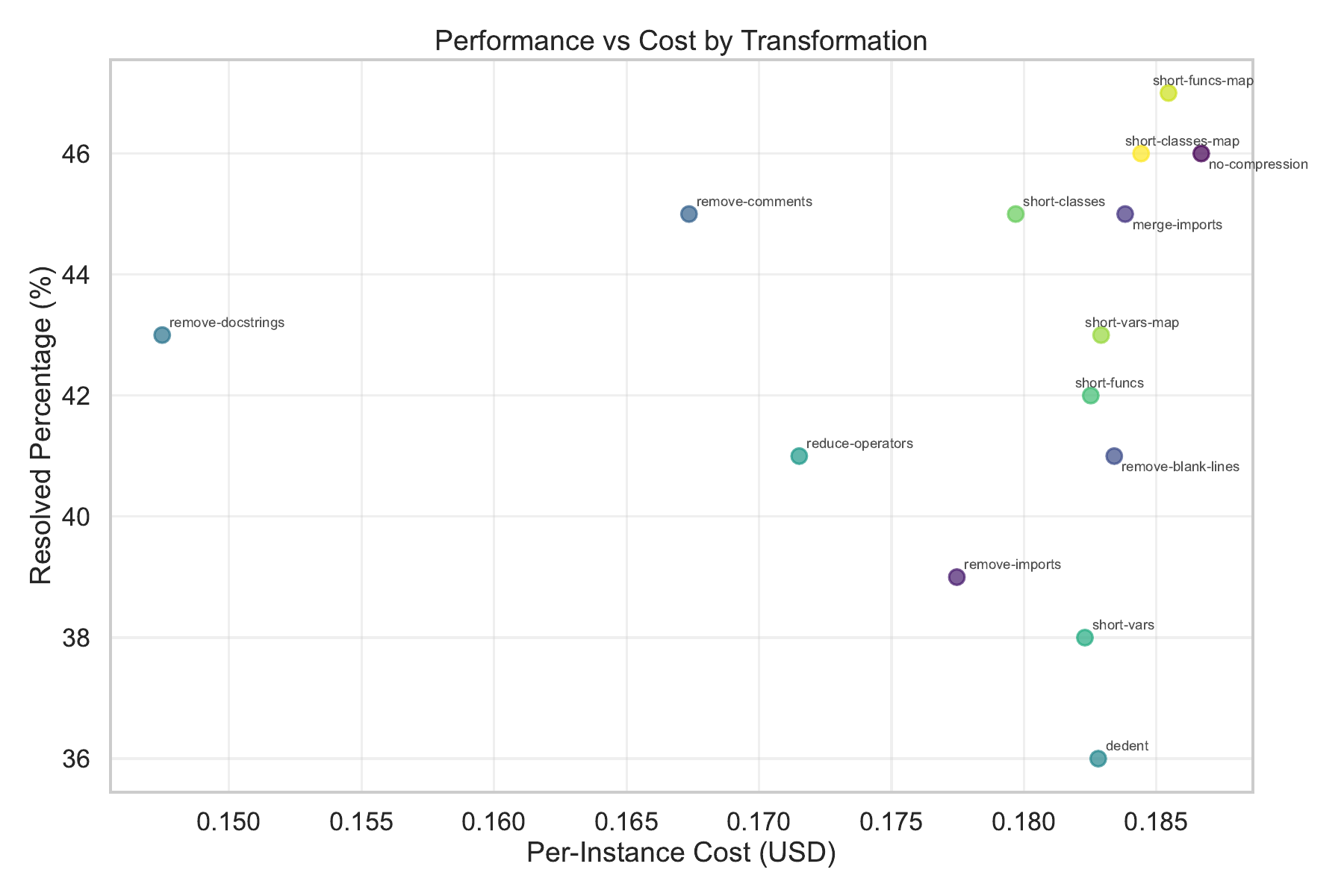}
    \caption{Performance vs. per-issue cost by transformation with GPT-4.1 on a 100-sample SWE-bench Verified subset.}
    \label{fig:performance-cost-individual}
\end{figure*}

In this ablation, we examine the effect of applying single transformations individually on cost and resolution rate. The results are summarized in \Cref{tab:single-transformations} and visualized in \Cref{fig:performance-cost-individual}. The most effective transformations in terms of token reductions are remove-comments and remove-docstrings, which decrease the average input size from 87{,}325 in the no-minification baseline to 77{,}849 and 68{,}202, respectively. These reductions translate directly into substantial per-instance cost savings, with remove-docstrings achieving the lowest overall cost at 0.1475~USD. The next most impactful transformation is reduce-operators (i.e., removing whitespace around operators), which lowers the average input tokens to 80{,}132 and the cost to 0.1715~USD. Other transformations, such as shortening identifiers or merging imports, yield smaller but still measurable reductions. In terms of performance, most transformations lead to relatively small decreases in resolution rate (typically below 5\% relative to the baseline). The most pronounced degradation occurs with dedent, which reduces the resolution rate to 36\%, corresponding to an 8\% drop compared to no minification.

\begin{table*}[t]
\centering
\begin{tabular}{lrrrr}
\hline
Transformation & Resolved (\%) & Input Tokens & Output Tokens & Cost (USD) \\
\hline
no-minification            & 46.00 & 87{,}325 & 1505 & 0.1867 \\
remove-imports             & 39.00 & 82{,}997 & 1434 & 0.1775 \\
merge-imports              & 45.00 & 86{,}034 & 1469 & 0.1838 \\
remove-blank-lines         & 41.00 & 85{,}190 & 1628 & 0.1834 \\
remove-comments            & 45.00 & 77{,}849 & 1457 & 0.1674 \\
remove-docstrings          & 43.00 & 68{,}202 & 1384 & 0.1475 \\
dedent                     & 36.00 & 84{,}754 & 1662 & 0.1828 \\
reduce-operators           & 41.00 & 80{,}132 & 1406 & 0.1715 \\
short-vars                 & 38.00 & 84{,}787 & 1590 & 0.1823 \\
short-funcs                & 42.00 & 85{,}070 & 1547 & 0.1825 \\
short-classes              & 45.00 & 83{,}423 & 1605 & 0.1797 \\
short-vars-map             & 43.00 & 85{,}260 & 1549 & 0.1829 \\
short-funcs-map            & 47.00 & 86{,}236 & 1623 & 0.1855 \\
short-classes-map          & 46.00 & 86{,}398 & 1453 & 0.1844 \\
\hline
\end{tabular}
\caption{Single transformation ablation results with GPT-4.1 on a 100-sample SWE-bench Verified subset.}
\label{tab:single-transformations}
\end{table*}

\subsubsection{Multiple Transformations}\label{ablation:multiple-transformations}

\begin{figure*}
    \centering
    \includegraphics[width=0.8\textwidth]{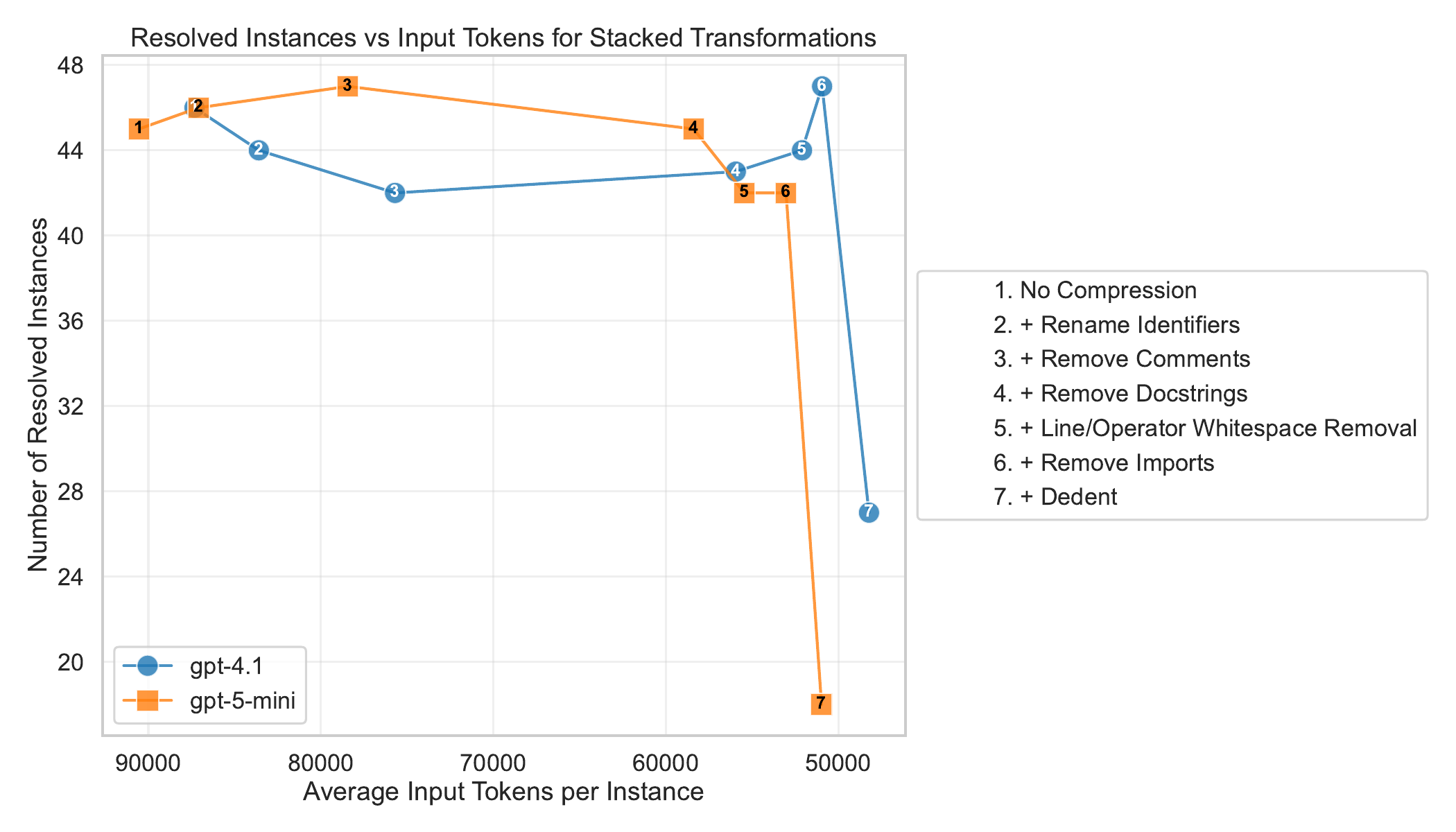}
    \caption{Resolved instances vs. average input tokens with stacked code minification transformations for GPT-4.1 and GPT-5-mini on a 100-sample SWE-bench Verified subset. Transformations are applied incrementally in the order listed in the legend.}
    \label{fig:performance_input_tokens_scatter_stacked}
\end{figure*}

\Cref{fig:performance_input_tokens_scatter_stacked} illustrates the effect of stacking multiple transformations for GPT-4.1 and GPT-5-mini. Transformations are applied in ascending order of their individual performance impact. Similar operations are grouped to reduce experiment cost. For instance, the renaming identifiers step adds variable, function, and class renaming. We also select remove-imports over merge-imports, as it has a higher potential for reducing token count. The plot should be interpreted from left to right, where each point represents the cumulative application of all preceding transformations plus the one indicated by the label. We find that stacking transformations consistently reduces per-instance cost while maintaining performance close to the baseline. The primary exception is again the dedent transformation, which reduces the resolution rate to 27\% compared to 47\% in the no-minification baseline. Apart from this case, the stacked transformations demonstrate that substantial reductions in token usage can be achieved with only modest losses in resolution performance. Across the two base LLMs, the overall trend seems consistent. GPT-5-mini achieves slightly higher performance with fewer transformations but degrades more rapidly under stronger minification. It also exhibits fewer input tokens on average at equivalent transformation stages. Since these values reflect repair-stage inputs, the difference likely arises from variations in file selection during ranking. However, we did not analyze this aspect further.

\subsection{Failure Analysis}

Among the evaluated transformations, most yield only minor performance degradation when applied individually or in stacked form. The main exception is the \emph{dedent} transformation, which produces a substantially larger decline in resolved instances. To understand this effect, we analyze instance-level outcomes and examine concrete failure cases.

\begin{figure*}[ht]
  \centering
  \includegraphics[width=\textwidth]{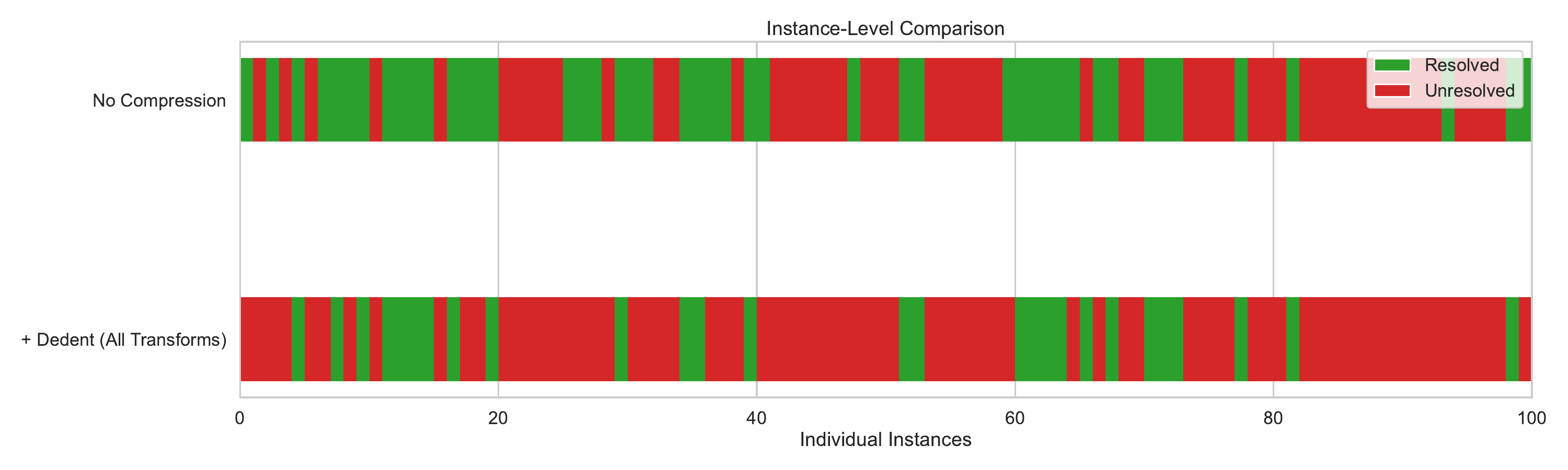}
  \caption{Comparing instance-level resolutions on SWE-bench Verified when no transformations (top) and all transformations (bottom) are used.}
  \label{fig:instance-level-dedent}
\end{figure*}

\Cref{fig:instance-level-dedent} compares the resolutions obtained with the unmodified baseline to those obtained when dedent is included. Each point corresponds to a benchmark instance; green denotes a successful resolution and red an unresolved case. The resolved instances under dedent form a strict subset of those resolved by the baseline, indicating that the observed degradation is unlikely to be caused by stochastic variability. Inspection of evaluation logs reveals that several failures are caused by syntactically invalid patches generated under dedentation. Although the repair prompt explicitly instructs the model to output SEARCH and REPLACE blocks using a four-space indentation style, the model occasionally fails to restore this indentation. As a result, the applied patches produce malformed Python code that triggers immediate test failures.

A representative example is the instance \texttt{django-12155}. The generated patch applies cleanly but yields an invalid function body, shown below:

\begin{lstlisting}[
  language=diff,
  breaklines=true,
  breakatwhitespace=false,
  columns=fullflexible,
  keepspaces=true
]
diff --git a/django/contrib/admindocs/utils.py b/django/contrib/admindocs/utils.py
index 4c0e7e2a56..2c5a41aae9 100644
--- a/django/contrib/admindocs/utils.py
+++ b/django/contrib/admindocs/utils.py
@@ -33,9 +33,9 @@ def trim_docstring(docstring):
     if not docstring or not docstring.strip():
         return ''
     # Convert tabs to spaces and split into lines
-    lines = docstring.expandtabs().splitlines()
-    indent = min(len(line) - len(line.lstrip()) for line in lines if line.lstrip())
-    trimmed = [lines[0].lstrip()] + [line[indent:].rstrip() for line in lines[1:]]
+    lines=docstring.expandtabs().splitlines()
+ indent=min(len(line)-len(line.lstrip())for line in lines[1:] if line.lstrip())
+ trimmed=[lines[0].lstrip()]+[line[indent:].rstrip()for line in lines[1:]]
     return "\n".join(trimmed).strip()
\end{lstlisting}

Lines originally nested inside a function lost their indentation hierarchy, producing the following Python error:
\begin{lstlisting}
IndentationError: unindent does not match any outer indentation level
\end{lstlisting}

This failure mode suggests that the dedent transformation primarily degrades performance by inducing syntactic errors rather than by removing semantically relevant information. A post-processing step that normalizes indentation in generated REPLACE blocks would likely mitigate these issues and bring dedent performance closer to that of other transformations.

\section{Limitations \& Future Work}

While the results demonstrate the effectiveness of token-reduction transformations for code-intensive repair tasks, several limitations constrain the precision and generality of the findings. The following section discusses these limitations and outlines directions for future research.

\subsection{Stochastic Variability in Model Behavior}

A primary limitation of this study arises from the stochastic nature of LLMs. Even under identical configurations, we observe variations in resolution rates across repair runs. Furthermore, the set of files selected during repair differs between runs, directly affecting the measured input and output token counts. Consequently, exact numerical results should be interpreted with caution, particularly when differences are small. Nonetheless, we argue that the general trends observed in our experiments remain valid and provide meaningful evidence of the effectiveness of the proposed transformations. Future work could mitigate these sources of variability through several strategies, for instance, by fixing the file selection process during repair and performing multiple runs per configuration to reduce variability.

\subsection{Agent-Specific Design Bias}

Another limitation of this study is that all experiments were conducted using a single agent that is particularly heavy on source code input. Consequently, the observed effects may not generalize to systems in which code ingestion constitutes a smaller proportion of the total context. Nonetheless, we hypothesize that source code processing represents a common bottleneck across a broad range of software engineering agents. When addressing issues, agents typically must read and interpret at least the directly relevant portions of the codebase to accurately localize and reason about the problem, which is an inherently challenging and context-intensive process.

\subsection{Validity of Generated Patches}

Recent work~\cite{wang2025plausiblepatchescorrect} has shown that automatically generated patches may produce behavior that satisfies test cases but diverges from developer intent. Specifically, LLM-based repair agents can generate plausible fixes that pass all tests while failing to resolve the root cause of the underlying issue or introducing unintended side effects. This study does not assess the semantic validity of generated patches beyond their test outcomes. Consequently, some patches classified as successful may not represent truly correct repairs. Future work could incorporate additional validation steps, such as human evaluation, to assess the quality and correctness of the produced repairs.

\subsection{Formatting Artifacts in Generated Diffs}\label{sec:formatting-artifacts}

Another limitation of the current implementation concerns the handling of code formatting during the repair process. In our setup, the LLM operates on minified code during the repair stage, producing minified SEARCH and REPLACE blocks that define the edits. The minified \texttt{SEARCH} block is then matched against the original (unminified) code using a search procedure that is agnostic to formatting changes introduced by minification. The corresponding \texttt{REPLACE} block is subsequently applied to the matched region in the original source. However, the transformations within the \texttt{REPLACE} block itself are not reversed, which can result in diffs containing unintended formatting modifications. An example of such an artifact is shown in \Cref{lst:formatting_diff}, where spacing around operators is removed as a side effect of minification (in addition to the intended import addition):

\begin{lstlisting}[language=diff, caption={Formatting artifacts introduced during patch application.}, label={lst:formatting_diff}, breaklines=true]
-from .util import _str_to_num, _is_int, translate, _words_group
+from.util import _str_to_num,_is_int,translate,_words_group,decode_ascii
\end{lstlisting}

This issue can be addressed with additional engineering effort, for instance, by applying a formatter that enforces the original code's style rules or by maintaining source maps that allow reversal of the applied transformations. We currently implement such reversibility only for the renaming transformations, as identifier renaming would otherwise break the test suite.

\section{Conclusion}

Using our re-implementation of the \emph{DirectSolve} \emph{state-in-context} agent, we reproduce the main performance trends reported by Jiang et al.~\cite{jiang2025puttingcontextsimplifyingagents} on SWE-bench Verified. Whereas their best DirectSolve configuration achieves a 50.8\% pass@1 rate with Gemini-2.5-Pro, our cost-controlled variant attains comparable pass@1 performance on the same benchmark using GPT-5-mini. This indicates that our implementation reproduces the main findings of the original paper.

Furthermore, empirical evaluation on SWE-bench Verified demonstrates that code minification reduces average input token usage by approximately 42\% compared to non-minified inputs, while decreasing the resolution rate by 12\%. This highlights an explicit trade-off between efficiency and effectiveness, with substantial reductions in context size accompanied by a measurable decrease in task performance. In addition, we evaluate the consistency of token savings and performance between models, repositories, and issue difficulty. GPT-4.1 and GPT-5-mini show similar behavior, with dedent as the primary outlier, causing a pronounced performance drop. The token reductions and performance effects also remain consistent across difficulty levels and are largely stable across repositories, indicating that the proposed transformations generalize well across different conditions.

\bibliography{bib}

\end{document}